\documentclass[twocolumn]{webofc}
\usepackage[varg]{txfonts} 

\usepackage[varg]{txfonts}   

\usepackage{amsmath}   
\usepackage{graphicx}  
\usepackage{verbatim}  
\usepackage{color}     
\usepackage{hyperref}  
\usepackage{bm}
\usepackage{ulem}
\usepackage{lineno}
\usepackage{booktabs}
\usepackage{dcolumn}

\begin{document}
\title{Two Pion Photo- and Electroproduction with CLAS}
%
%

\author{\firstname{Victor I.} \lastname{Mokeev for the CLAS Collaboration}\inst{1}\fnsep\thanks{\email{mokeev@jlab.org}} 
}

\institute{Jefferson Laboratory, 12000 Jefferson Ave., Newport News VA, 23606, USA 
          }

\abstract{
Exclusive $\pi^+\pi^-p$ photo- and electroproduction data from CLAS have considerably extended the information on the spectrum and structure of nucleon resonances. The data from the $\pi^+\pi^-p$ and $N\pi$ channels have provided results on the electrocouplings of most resonances in the mass region up to 1.8 GeV and at photon virtualities up to 5.0 GeV$^2$. The recent CLAS data on $\pi^+\pi^-p$ photoproduction have improved knowledge on the photocouplings of nucleon resonances in the mass range of 1.6 GeV $<$ $M_{N^*}$ $<$ 2.0 GeV and on their decays to the $\pi \Delta$ and $\rho p$ final hadron states. 
Analyses of the combined $\pi^+\pi^-p$ photo- and electroproduction data have revealed evidence for the candidate-state $N'(1720)3/2^+$. The new results on the nucleon resonance spectrum, electroexcitation amplitudes from analysis of the CLAS $\pi^+\pi^-p$ photo- and electroproduction data, and their impact on the exploration of strong QCD are presented. 
}

\maketitle

\section{Introduction}
\label{intro}
Studies of exclusive $\pi^+\pi^-p$ photo- and electroproduction off protons represent an effective tool for the exploration of the spectrum and structure of excited nucleon states ($N^*$). This exclusive channel is sensitive to the contributions of most $N^*$, which allows for the determination of the nucleon resonance photo- and electrocouplings ($\gamma_{r,v}pN^*$) at different photon virtualities \cite{Mokeev:2018zxt}. This provides insight into the structure of most nucleon resonances with their distinctively different structural features. The exploration of different resonance electrocouplings has revealed excited nucleon state structure as a complex interplay between the inner core of three dressed quarks and an external meson-baryon cloud \cite{Bu19n,Craig19n,Burkert:2019bhp,Aznauryan:2018okk,Kamano:2018sfb}. The continuum QCD analyses of the ground nucleon elastic form factors and
$\gamma_{v}pN^*$ electrocouplings of different $N^*$ starting from the QCD Lagrangian already shed light on the strong QCD dynamics underlying the generation of the dominant part of hadron mass \cite{Craig19n,Segovia:2014aza,Segovia:2015hra,Mokeev:2015lda}. Essential progress has been achieved in the developments of the quark models for the description of the structure of many excited nucleon states \cite{Aznauryan:2018okk,Aznauryan:2012ec,Ramalho:2018wal,Gil19n,Obukhovsky:2013fpa,Giannini:2015zia,Santo19n}. Extension of the results on the $\gamma_{r,v}pN^*$  electrocouplings toward larger photon virtualities and for the majority excited nucleon states will offer a unique opportunity to explore many facets of strong QCD in the generation of different nucleon resonances.
 
The recent CLAS data on exclusive $\pi^+\pi^-p$ photoproduction \cite{Golovatch:2018hjk} allow us to determine the photocouplings of almost all well- established resonances in the mass range from 1.6 GeV to 2.0 GeV for the first time from  this exclusive channel. Analysis of the recent CLAS data \cite{Isupov:2017lnd,Trivedi:2018rgo} on $\pi^+\pi^-p$ electroproduction at 1.4 GeV $<$ $W$ $<$ 2.0 GeV and 2.0 GeV$^2$ $<$ $Q^2$ $<$ 5.0 GeV$^2$ demonstrated promising prospects for the extraction of the $\gamma_{r,v}pN^*$  electrocouplings of most resonances in the above mentioned kinematics area. The combined studies of $\pi^+\pi^-p$ photo- and electroproduction have revealed evidence for a new candidate-state $N'(1720)3/2^+$ \cite{Mokeev:2018zxt}. Further development of the JM reaction model for the extraction of the $\gamma_{r,v}pN^*$  photo-/electrocouplings from the $\pi^+\pi^-p$ exclusive channel based on analysis of the recent CLAS data \cite{Isupov:2017lnd,Trivedi:2018rgo}, the new results on the $\gamma_{r,v}pN^*$  photo-/electrocouplings and their impact on strong QCD exploration and the observation of the new $N'(1720)3/2^+$ state will be presented in this proceeding.

\section{Experimental Studies of $\pi^+\pi^-p$ Photo-/Electroproduction and the Reaction Model}
\label{data_anal}
The combination of the CEBAF CW beam and the CLAS detector with almost 4$\pi$ acceptance provides a unique opportunity for the exploration of the $\gamma_{r,v} p \rightarrow \pi^+\pi^-p'$ reactions with almost complete coverage of the final state phase space. This is of particular importance for the extraction of the  photo- and $\gamma_{v}pN^*$ electrocouplings, as well as their hadronic decay widths into the $\pi\Delta$ and $\rho p$ final states. 
Table~\ref{clas_data} summarizes the available CLAS experimental data on exclusive $\pi^+\pi^-p$ photo- and electroproduction. The new $\pi^+\pi^-p$ photoproduction off 
proton data \cite{Golovatch:2018hjk} were obtained with the CLAS detector for the first time in terms of nine independent 1-fold differential and 
fully integrated $\pi^+\pi^-p$ cross sections in the center of mass energies $W$ from 1.6~GeV to 2.0~GeV. Overall, $\approx$400 million $\pi^+\pi^-p$ events were selected, exceeding by a factor of $\sim$50 the statistics previously collected in this channel. The $\pi^+\pi^- p$ electroproduction data from CLAS~\cite{Isupov:2017lnd,Trivedi:2018rgo} have extended the information on the nine independent single-differential and fully integrated cross sections binned in $W$ and $Q^2$ in the mass range $W < 2.0$~GeV towards large photon virtualities of 2.0~GeV$^2 < Q^2 < 5.0$~GeV$^2$. The first results on the structure functions $R_{TT}$ and $R_{TL}$ for the $\gamma_{v} p \rightarrow \pi^+\pi^-p'$ channel have become available \cite{Trivedi:2018rgo}. The measured observables can be found in the CLAS Physics Data Base \cite{clasdb}.

\begin{table} []
\begin{center}
\caption{The kinematic areas covered in the CLAS measurements of exclusive $\pi^+\pi^-p$ photo- and electroproduction 
nine differential and fully integrated cross sections. Ref~\cite{Trivedi:2018rgo} additionally includes the structure functions. }
\label{clas_data}      
\begin{tabular}{|c|c|c|}
\hline
$W$-ranges, & $Q^2$-ranges, & Refs. \\
GeV       & GeV$^2$       &        \\\hline
1.6-2.0   & 0.            & \cite{Golovatch:2018hjk}  \\  \hline
1.3-1.57  & 0.2-0.6       & \cite{Fedotov:2008aa} \\  \hline
1.4-2.1   & 0.5-1.5       & \cite{Ri03} \\  \hline 
1.3-1.83  & 0.4-1.0       & \cite{Fedotov:2018oan} \\  \hline
1.4-2.0   & 2.0-5.0       & \cite{Isupov:2017lnd} \\  \hline
1.4-2.0   & 2.0-5.0       & \cite{Trivedi:2018rgo} \\  \hline
\end{tabular}
\end{center}
\end{table}

\begin{table}[]
\begin{center}
\caption{The values of $\chi^2/d.p.$ from the comparison between the measured and computed within the updated JM model nine one-fold 
differential $\gamma_{v}p \rightarrow \pi^+\pi^-p$ cross sections.}
\label{fit_chi2} 
\begin{tabular}{|c|c|c|c|c|c|} \hline
$Q^2$ bins& 2.0-2.4 & 2.4-3.0 & 3.0-3.5 & 3.5-4.2 & 4.2-5.0  \\
  GeV$^2$ &         &         &         &         &  \\  \hline
 $\chi^2/d.p.$         &   2.68  &  2.93   &  2.53   &   2.30   & 2.10 \\  \hline 
\end{tabular}

\end{center}
\end{table}

\begin{figure}[] 
\begin{center}
\includegraphics[width=5.5cm, height=6.0cm]{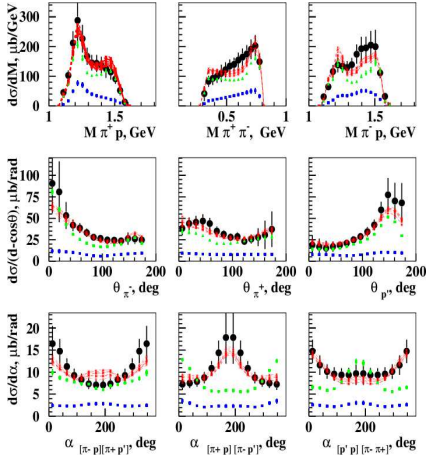}
\caption{(Color online) Representative
example of 1-fold differential cross sections (points with error bars) and the resonant/non-resonant contributions (blue/green bars) from
the fits (red curves) of the CLAS $\pi^+\pi^-p$ photoproduction data at $W$ from 1.73~GeV to 1.75~GeV.}
\label{res_backgr}
\end{center} 
\end{figure}

\begin{figure} 
\begin{center}
\includegraphics[width=7.cm, height=6.5cm]{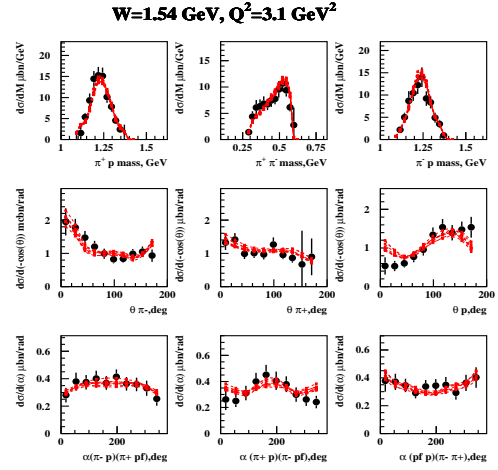}
\caption{(Color online) Data for $\pi^+\pi^-p$ electroproduction \cite{Trivedi:2018rgo} (black filled circles with error bars) and differential cross sections computed (red lines) within the updated JM model (see Section ~\ref{data_anal})  at 1.53 GeV $<$ $W$ $<$ 1.55 GeV and 3.0 GeV$^2$ $<$ $Q^2$ $<$ 3.5 GeV$^2$.}
\label{trivedi_fit}
\end{center} 
\end{figure}

\begin{figure*}
\begin{center}
\includegraphics[width=6.cm, height=7.cm]{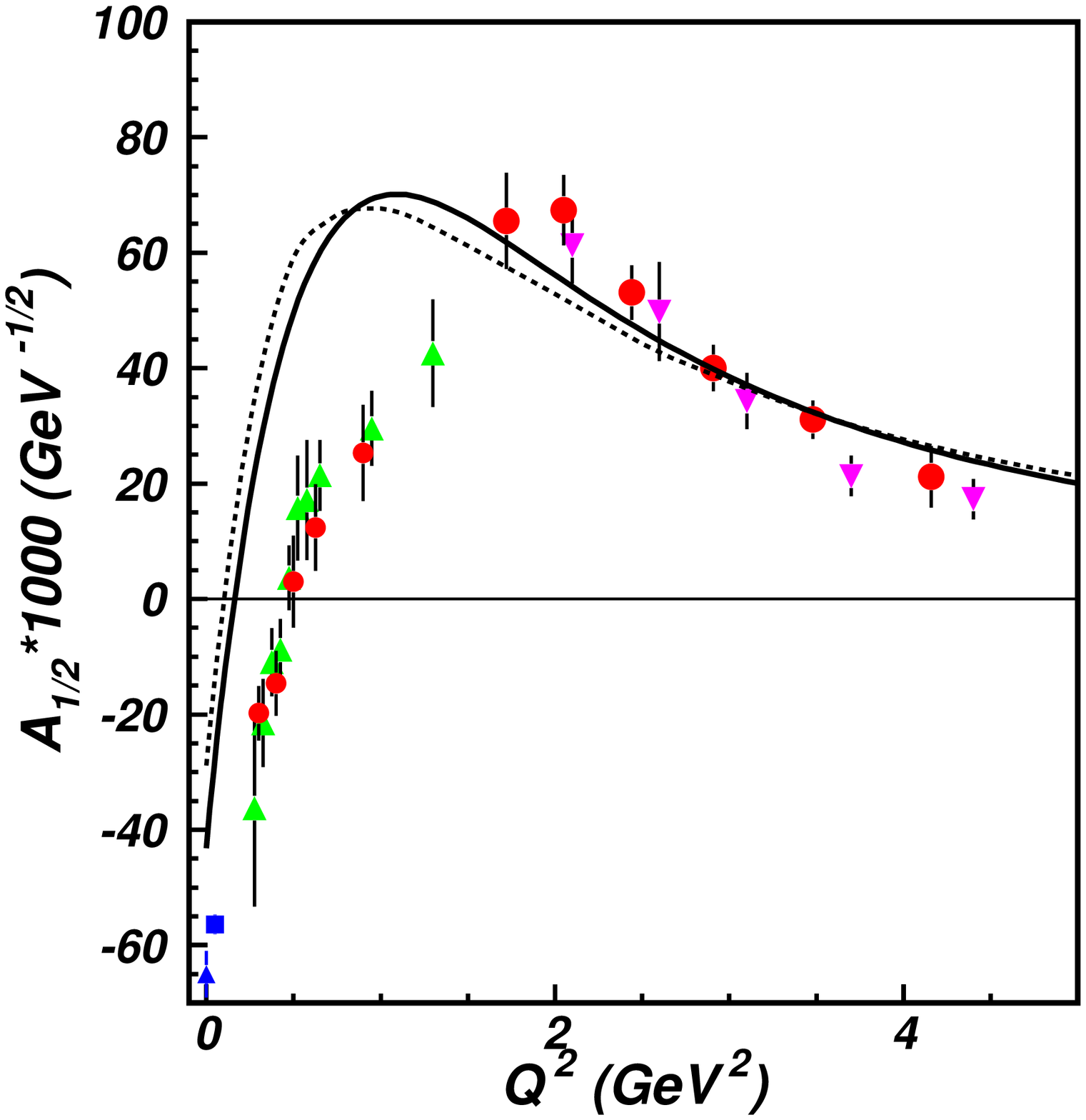}
\hspace{0.cm}
\includegraphics[width=6.cm, height=7.cm]{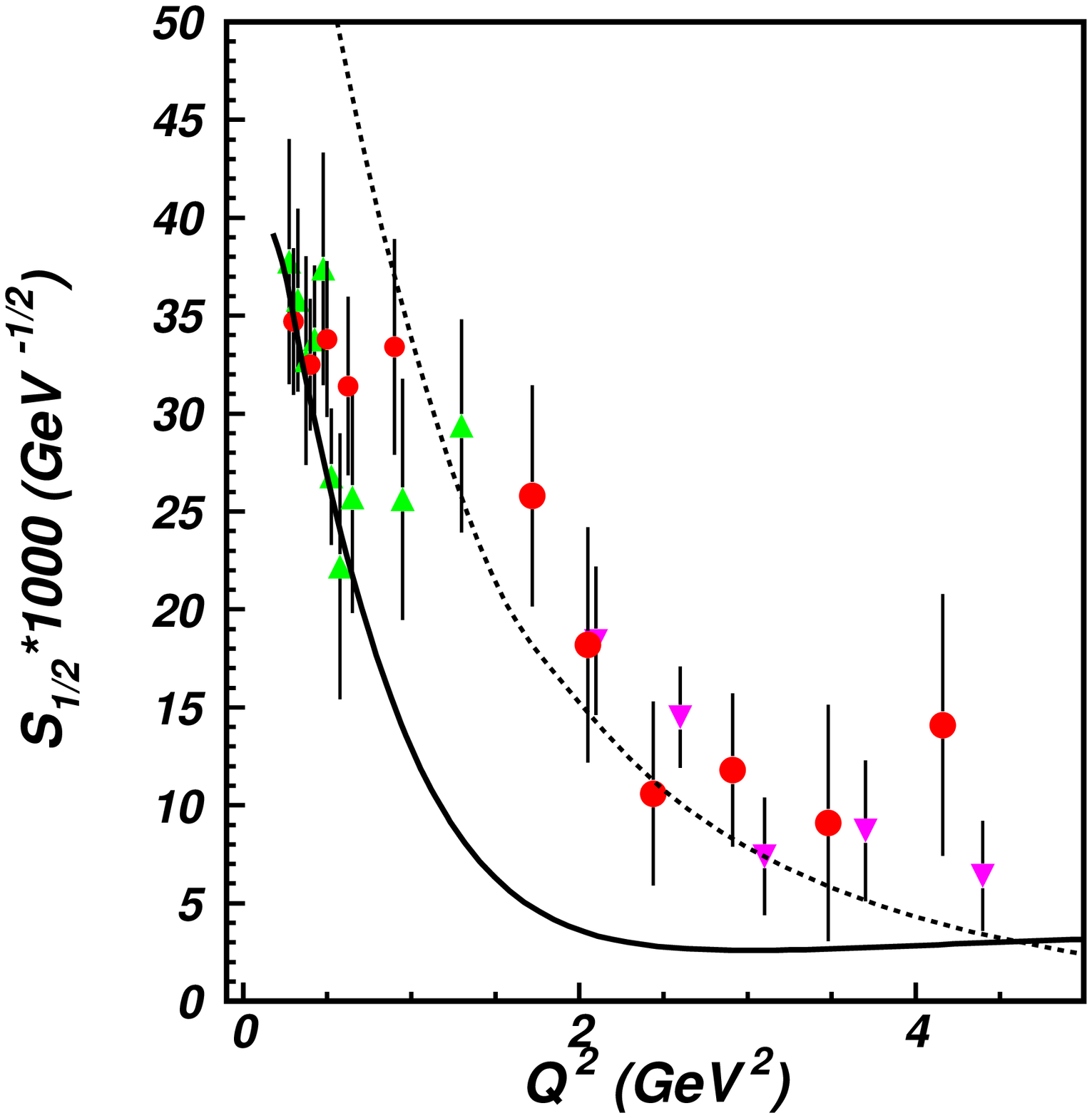}
\vspace*{0.3cm}
\caption{(Color online) CLAS results on the $N(1440)1/2^+$ electrocouplings from analysis of the recent data on $\pi^+\pi^-p$ electroproduction \cite{Isupov:2017lnd,Trivedi:2018rgo} (magenta filled triangles), from the analyses of $N\pi$ electroproduction \cite{Aznauryan:2009mx} (red filled circles), and from the previous CLAS data on $\pi^+\pi^-p$ electroproduction \cite{Mokeev:2015lda, Mokeev:2012vsa} (green filled circles). The results at the photon point are taken from the PDG \cite{Tanabashi:2018oca} (blue filled triangle) and  from analysis of the CLAS $N\pi$ photoproduction data \cite{Dug09} (blue filled square) The $N(1440)1/2^+$ electrocouplings computed within the DSE \cite{Segovia:2015hra} are shown by the solid lines. The results from the novel light front quark model \cite{Aznauryan:2012ec,Aznauryan:2018okk} are shown by the dashed lines.}
\label{p11elec}
\end{center} 
\end{figure*}

\begin{table*}[]
\begin {center}
\caption{Resonance photocouplings determined from analysis of the $\pi^+\pi^-p$ photoproduction data from this work in 
comparison with the previous results from the PDG average \cite{Tanabashi:2018oca} and from multichannel analysis \cite{Sokhoyan:2015fra}.}
\label{nstpar}
\begin{tabular}{ c  
                 D{,}{\pm}{-1}  
		 D{,}{\hspace{3pt} \text{--} \hspace{3pt}}{-1} 
                 D{,}{\pm}{-1}   
		 D{,}{\pm}{-1} 
		 D{,}{\hspace{3pt} \text{--} \hspace{3pt}}{-1}   
		 D{,}{\pm}{-1}}

\toprule
{}
& \multicolumn{1}{l}{$A_{1/2}\times 10^3$}
& \multicolumn{1}{l}{$A_{1/2}\times 10^3$}
& \multicolumn{1}{l}{$A_{1/2}\times 10^3$}
& \multicolumn{1}{l}{$A_{3/2}\times 10^3$}
& \multicolumn{1}{l}{$A_{3/2}\times 10^3$}
& \multicolumn{1}{l}{$A_{3/2}\times 10^3$}
\\
{Resonances}
& \multicolumn{1}{l}{from $\pi^+\pi^-p$}
& \multicolumn{1}{l}{PDG ranges}
& \multicolumn{1}{l}{multichannel}
& \multicolumn{1}{l}{from $\pi^+\pi^-p$}
& \multicolumn{1}{l}{PDG ranges}
& \multicolumn{1}{l}{multichannel}
\\
{}
& \multicolumn{1}{l}{GeV$^{-1/2}$}
& \multicolumn{1}{l}{GeV$^{-1/2}$}
& \multicolumn{1}{l}{analysis~\cite{Sokhoyan:2015fra}}
& \multicolumn{1}{l}{GeV$^{-1/2}$}
& \multicolumn{1}{l}{GeV$^{-1/2}$}
& \multicolumn{1}{l}{analysis~\cite{Sokhoyan:2015fra}}
\\
& \multicolumn{1}{l}{}
& \multicolumn{1}{l}{}
& \multicolumn{1}{l}{GeV$^{-1/2}$}
& \multicolumn{1}{l}{}
& \multicolumn{1}{l}{}
& \multicolumn{1}{l}{GeV$^{-1/2}$}
\\
\midrule
$\Delta(1620)1/2^-$    & 29.0,6.2    &	30 , 60     &   55,7      & 	        &  	        &	     \\
$N(1650)1/2^-$         & 60.5,7.7    &	35 , 55     &   32,6      & 	        &  	        &	     \\
$N(1680)5/2^+$         & -27.8,3.6   &  -18 ,{-5}   &  -15,2      & 128,11      &  130 , 140    &  136 ,  5  \\
$N(1720)3/2^+$         & 80.9,11.5   &  80 , 120    &   115,45    & -34.0,7.6   &  -48 , 135    &  135 , 40  \\
$\Delta(1700)3/2^-$    & 87.2,18.9   &  100 , 160   &   165,20    & 87.2,16.4   &   90 , 170    &  170 , 25  \\
$\Delta(1905)5/2^+$    & 19.0,7.6    &  17 , 27     &   25,5      & -43.2,17.3  &  -55 , {-35}  &  -50  , 5  \\
$\Delta(1950)7/2^+$    & -69.8,14.1  &  -75 , {-65} &   -67,5     & -118.1,19.3 &  -100 , {-80} &  -94 , 4   \\ 
\bottomrule	        	      
\end{tabular}
\end {center}
\end{table*}

\begin{figure*}[]
\begin{center}
\includegraphics[width=4.5cm, height=8.5cm]{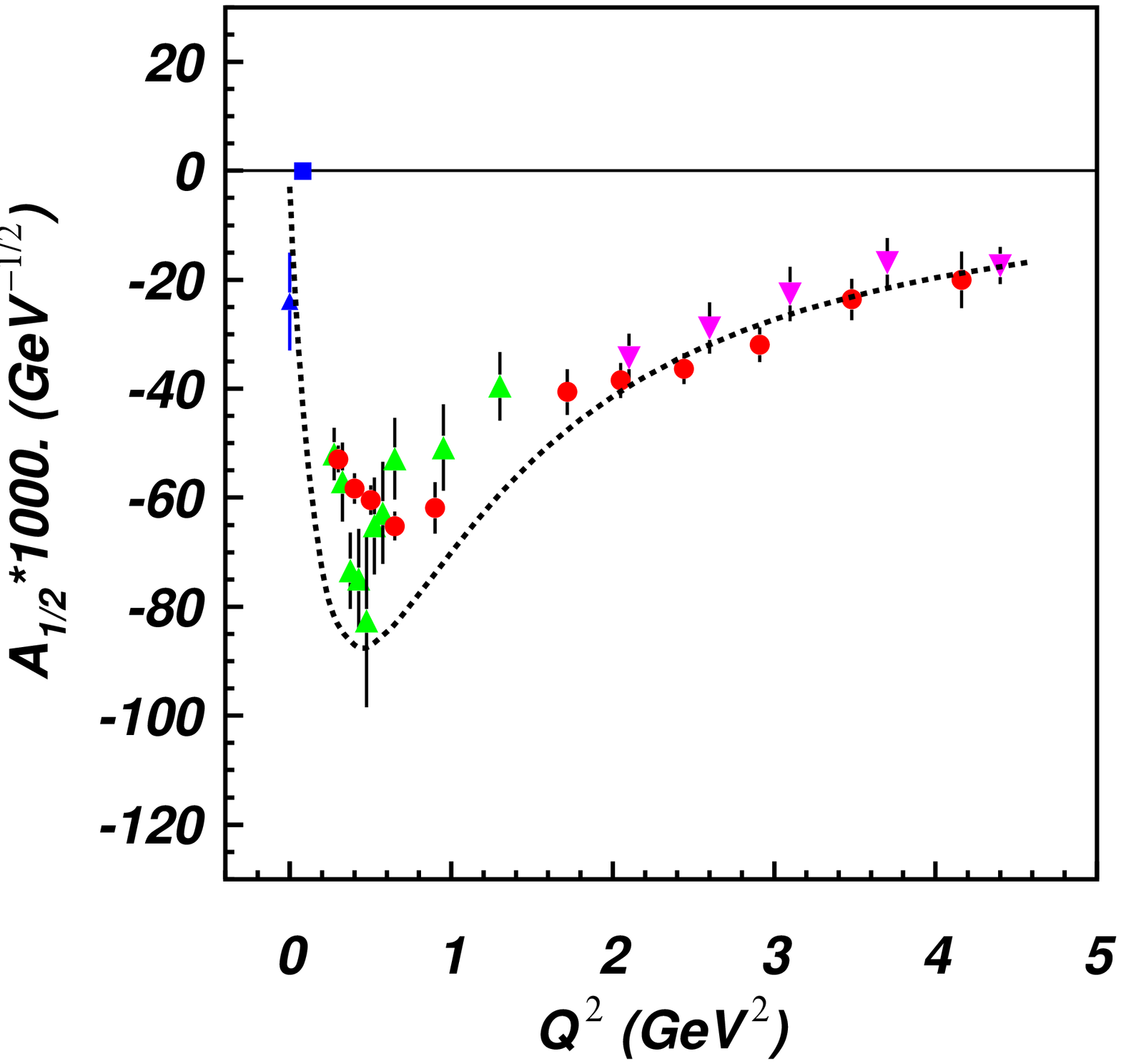}
\hspace{0.cm}
\includegraphics[width=4.5cm, height=8.5cm]{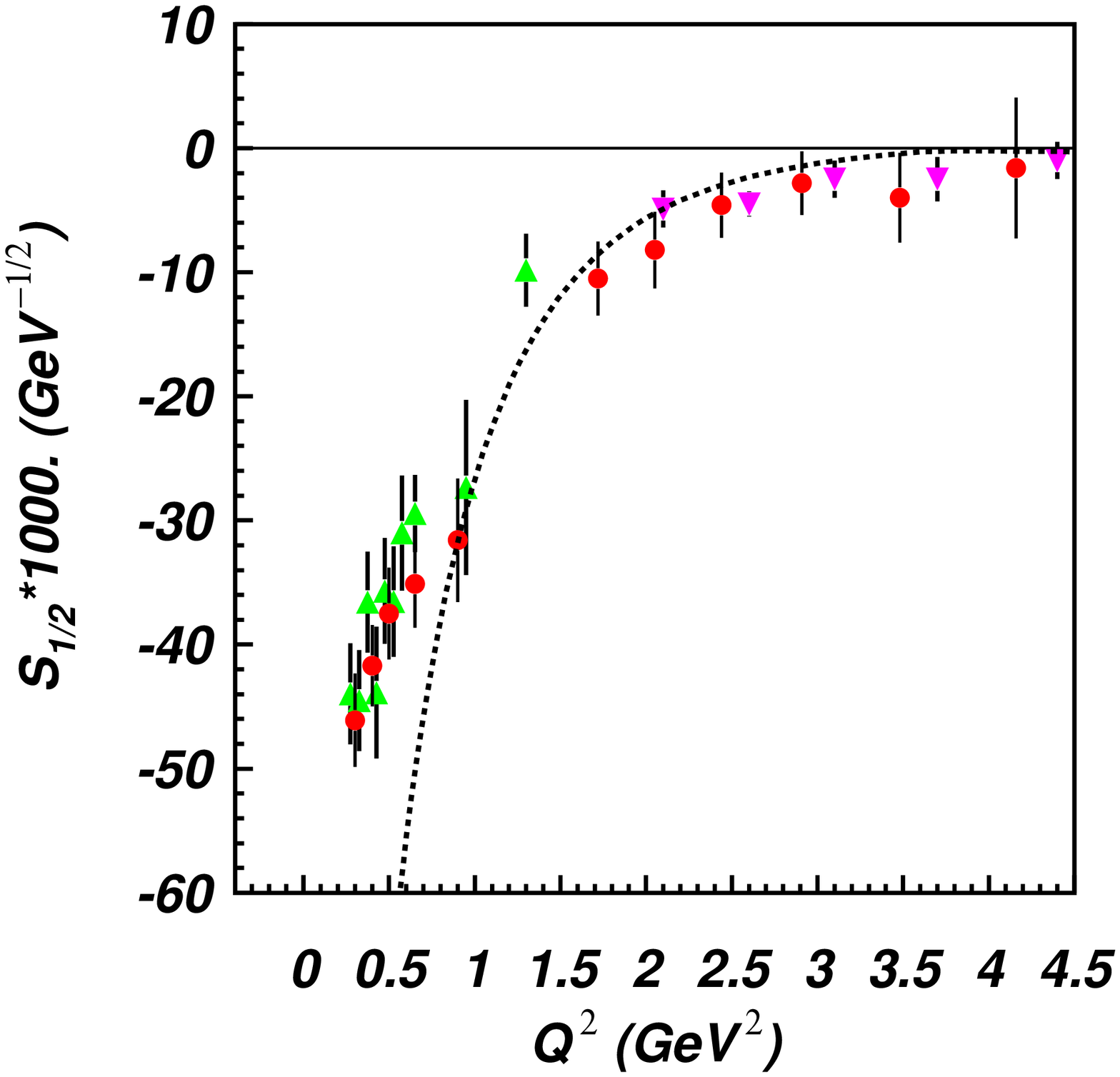}
\hspace{0.cm}
\includegraphics[width=4.5cm, height=8.5cm]{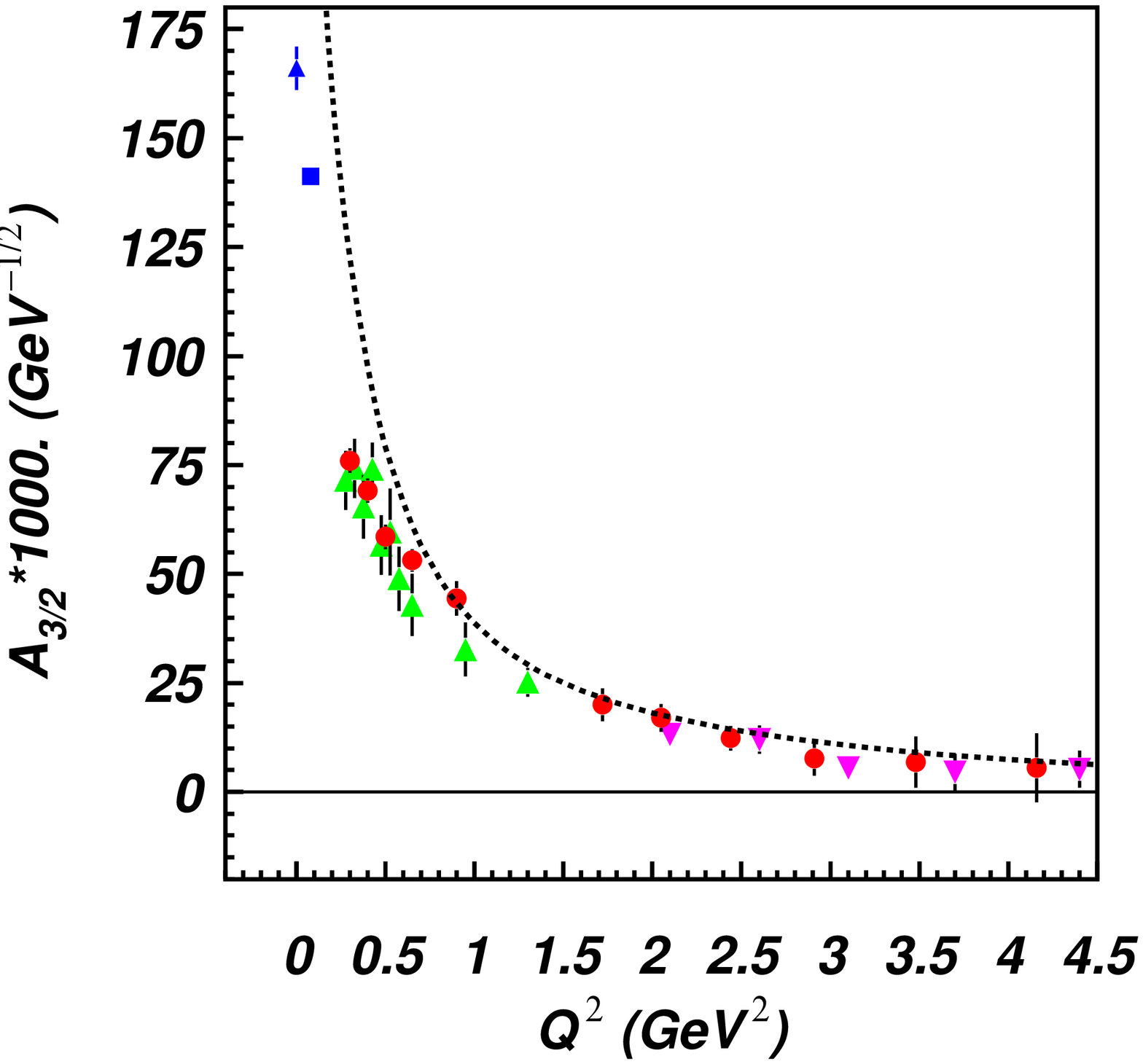}
\vspace*{0.2cm}
\caption{(Color online) CLAS results on the $N(1520)3/2^-$ electrocouplings from analysis of the recent data on $\pi^+\pi^-p$ electroproduction \cite{Isupov:2017lnd,Trivedi:2018rgo} (magenta filled triangles), from the analyses of $N\pi$ electroproduction \cite{Aznauryan:2009mx} (red filled circles), and from the previous CLAS data on $\pi^+\pi^-p$ electroproduction \cite{Mokeev:2015lda,Mokeev:2012vsa} (green filled circles). The results at the photon point are taken from PDG \cite{Tanabashi:2018oca} (blue filled triangles) and  from analysis of the CLAS $N\pi$ photoproduction data \cite{Dug09} (blue filled squares) The results from the novel light front quark model \cite{Aznauryan:2012ec,Aznauryan:2018okk} are shown by the dashed lines. }
\label{d13elec}
\end{center} 
\end{figure*}

\begin{figure*} []
\begin{center}
\includegraphics[width=6.3cm, height=7.cm]{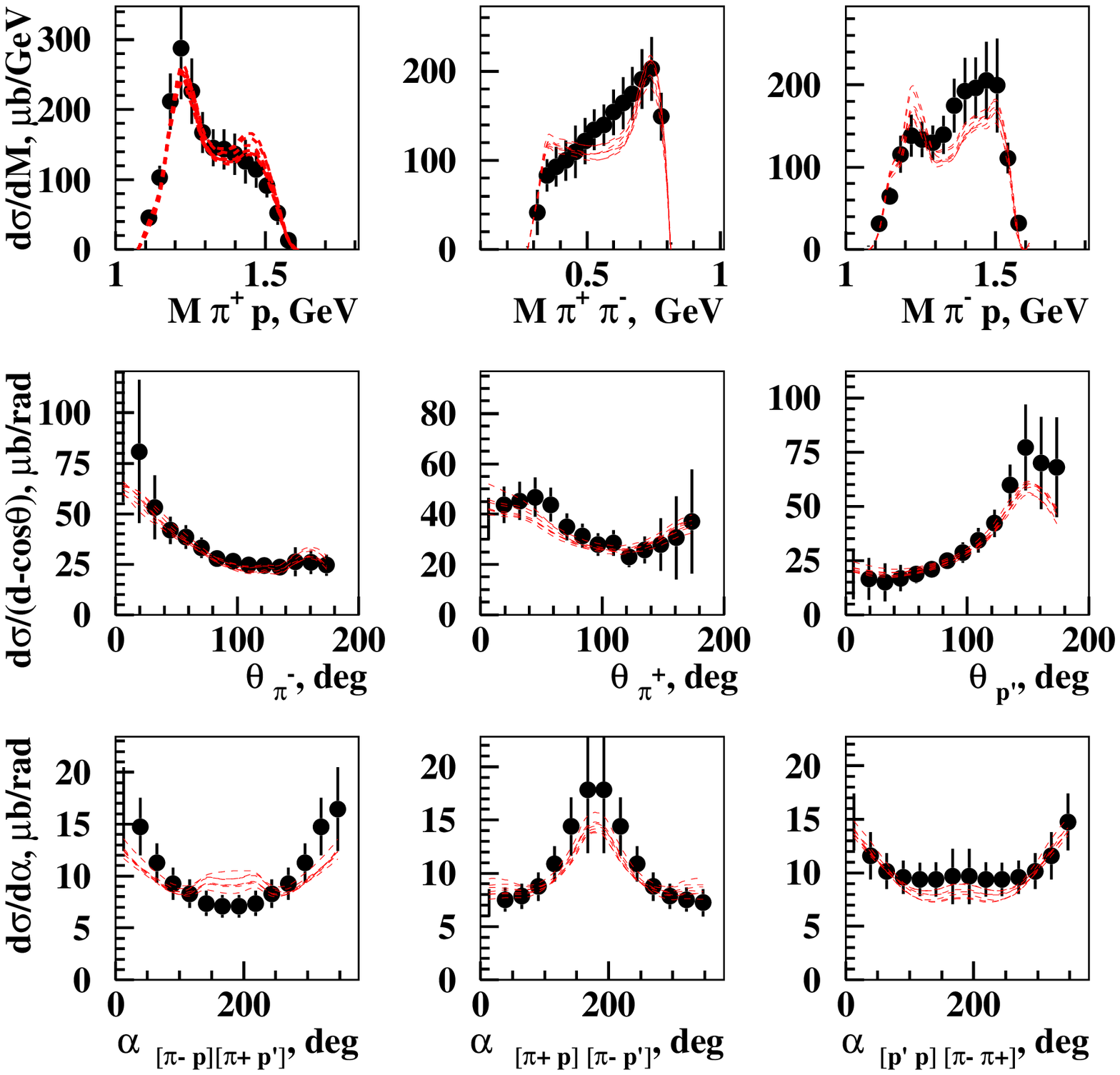}
\hspace{0.cm}
\includegraphics[width=6.3cm, height=7.cm]{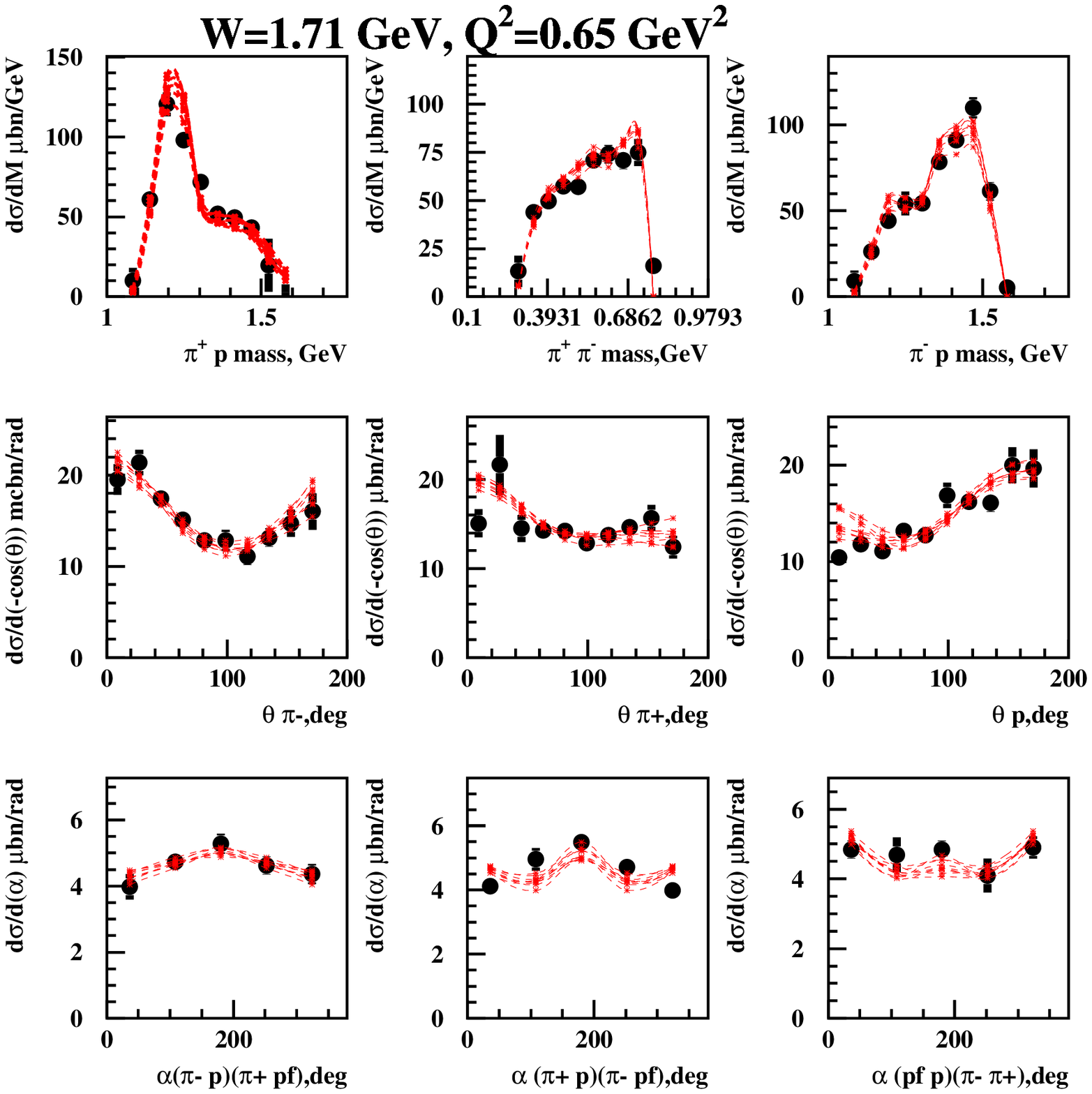}
\vspace*{0.2cm}
\caption{(Color online) Description (red curves) of the CLAS $\pi^+\pi^-p$ photoproduction data (black filled circles with error bars) at $W$ = 1.74 GeV and $Q^2$= 0. GeV$^2$ \cite{Golovatch:2018hjk} (left) and electroproduction data at $W$ = 1.74 GeV and $Q^2$= 0.65 GeV$^2$ \cite{Ri03} (right) achieved in the fit within the JM model after implementation of the candidate $N'(1720)3/2^+$ state.}
\label{missing_res_sect}
\end{center} 
\end{figure*}

\begin{figure*}[]
\begin{center}
\includegraphics[width=4.5cm, height=5.5cm]{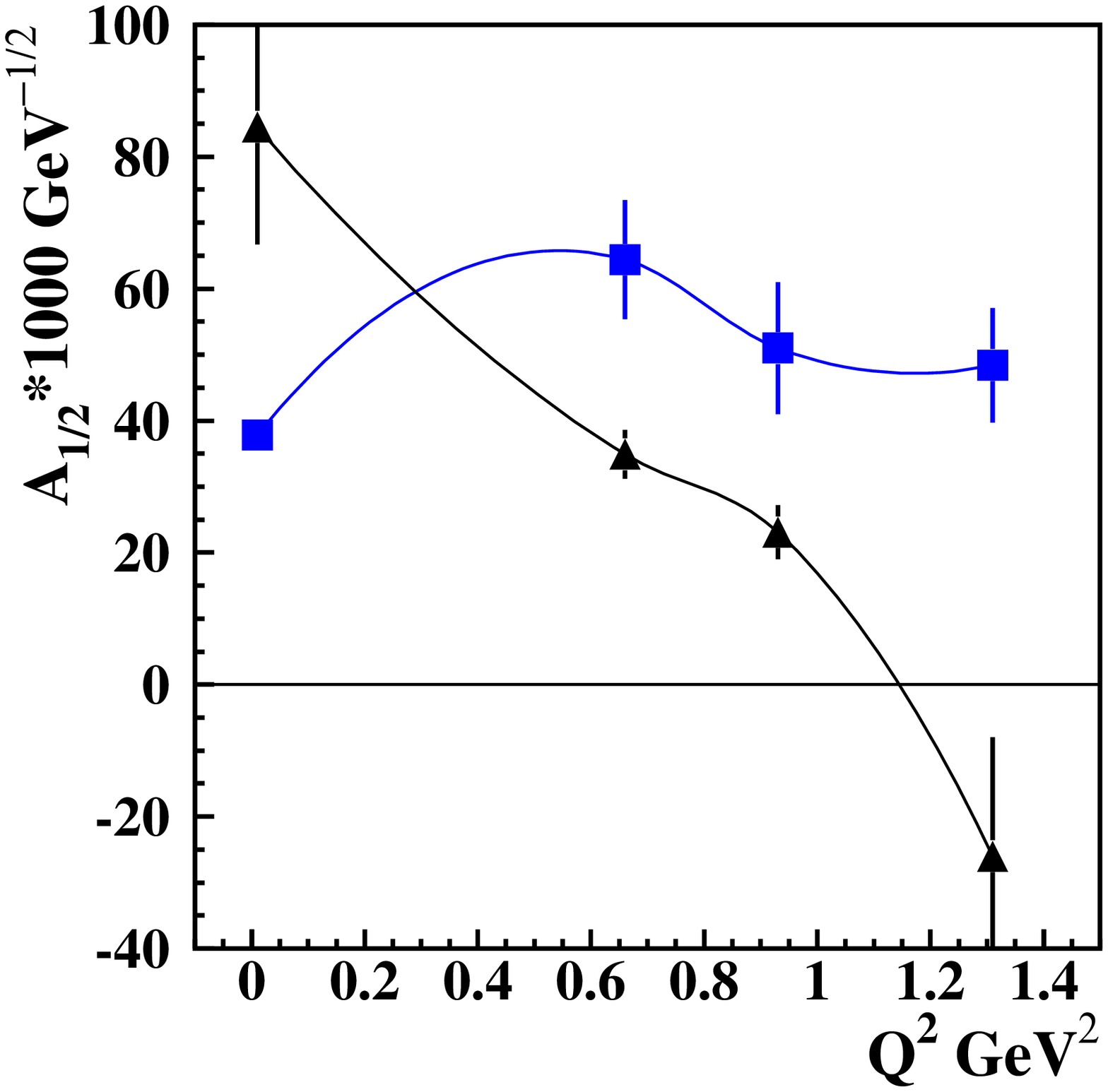}
\hspace{0.cm}
\includegraphics[width=4.5cm, height=5.5cm]{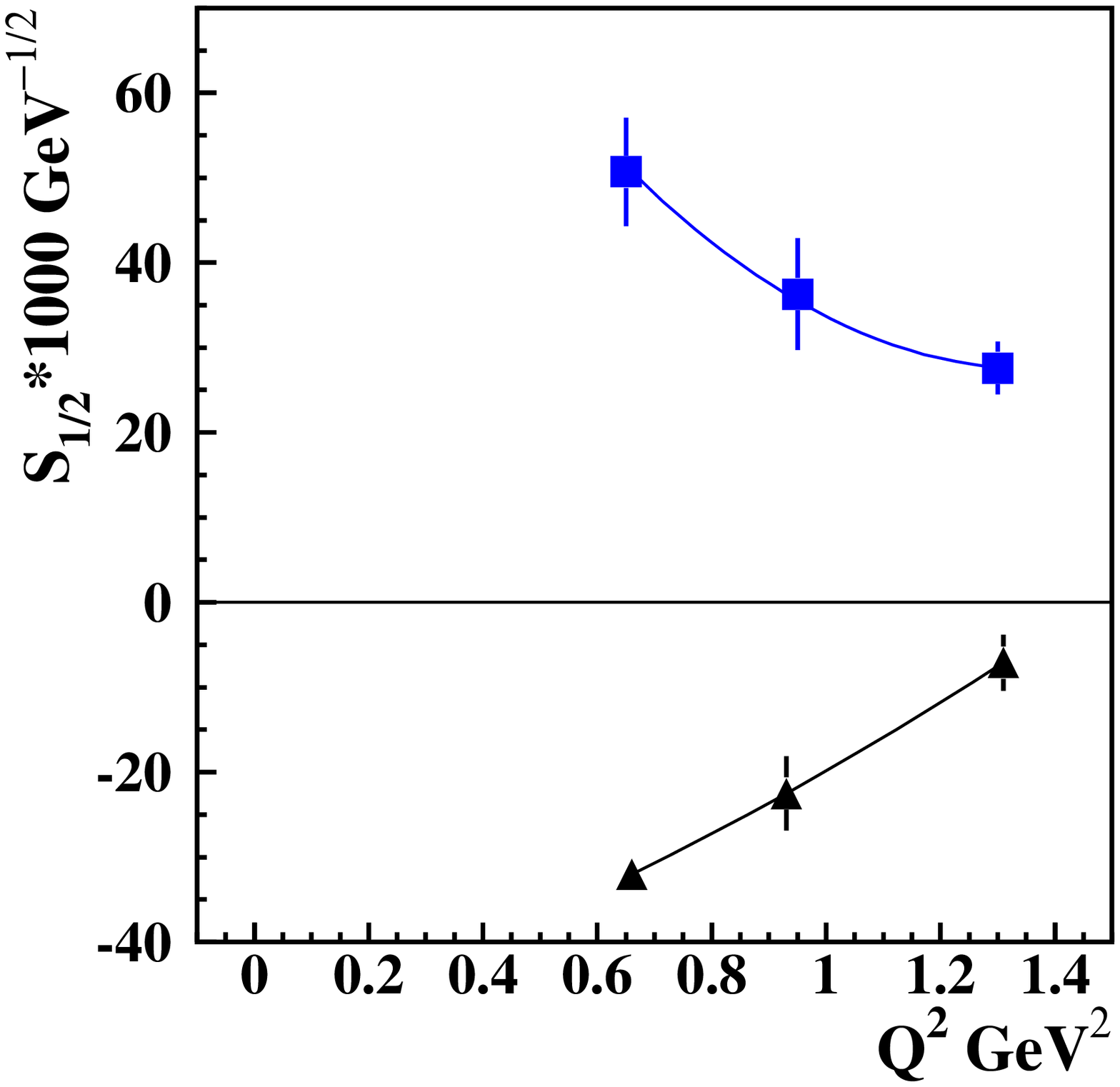}
\hspace{0.cm}
\includegraphics[width=4.5cm, height=5.5cm]{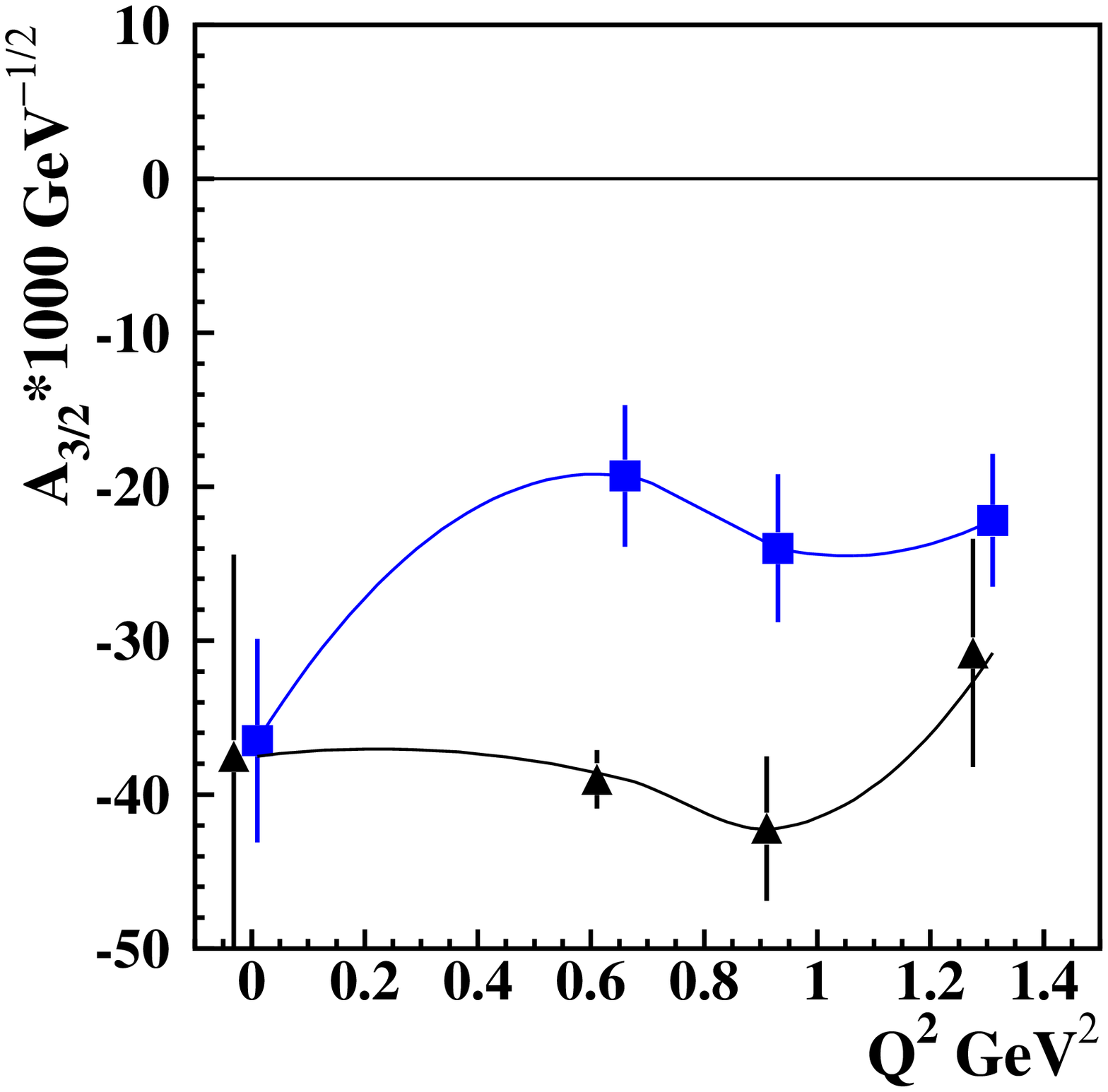}
\vspace*{0.1cm}
\caption{(Color online) Photo-/electrocouplings of the conventional $N(1720)3/2^+$ (filled triangles) and candidate $N'(1720)3/2^+$ (filled squares) states determined from the CLAS data on $\pi^+\pi^-p$ photo- and electroproduction \cite{Golovatch:2018hjk,Ri03} within the JM model.}
\label{miss_conv_elec}
\end{center} 
\end{figure*}

 The recent CLAS data  on the $\pi^+\pi^-p$ photo-/electroproduction cross sections and the first results on the exclusive structure functions for this channel \cite{Golovatch:2018hjk,Isupov:2017lnd,Trivedi:2018rgo} offer a promising prospect for the evaluation of the  photocouplings for the resonances with masses from 1.6 GeV to 2.0 GeV and $\gamma_{v}pN^*$ electrocouplings of most excited nucleon states in the mass range up to 2.0 GeV and photon virtualities $Q^2$ from 2.0 GeV$^2$ to 5.0 GeV$^2$. In preparation for the extraction of the $\gamma_{v}pN^*$ electrocouplings from these data, the JM reaction model \cite{Mokeev:2008iw,Mokeev:2012vsa,Mokeev:2015lda}, which was successfully used for the extraction of $\gamma_{v}pN^*$ electrocouplings from the $\pi^+\pi^-p$ electroproduction data at 0.2 GeV$^2$ $<$ $Q^2$ $<$ 1.5 GeV$^2$ \cite{Mokeev:2012vsa,Mokeev:2015lda}, was updated. The major focus of the recent CLAS data analysis is to establish all relevant mechanisms contributing to $\pi^+\pi^-p$ photoproduction in the kinematics area of 1.6 GeV $<$ $W$ $<$ 2.0 GeV  and in the kinematics area of 1.4 GeV $<$ $W$ $<$ 2.0 GeV and 2.0 GeV$^2$ $<$ $Q^2$ $<$ 5.0 GeV$^2$ in electroproduction from their manifestations in the kinematic dependencies of the measured differential cross sections and eventually to isolate the resonant contributions. The full $\gamma_{r,v} p \to \pi^+\pi^- p'$ amplitudes are described in the JM model as a superposition of the $\pi^-\Delta^{++}$, $\pi^+\Delta^0$, $\rho p$, $\pi^+ D_{13}^0(1520)$, and $\pi^+ F_{15}^0(1685)$ sub-channels with subsequent decays of the unstable hadrons to the final $\pi^+\pi^-p$ state and direct 2$\pi$ production mechanisms, where the final $\pi^+\pi^- p$ state comes about without going through the intermediate process of forming unstable hadron states.The JM model incorporates contributions from all well-established $N^*$ states of four-star PDG rating \cite{Tanabashi:2018oca} with the observed decays to the $\pi \Delta$ and $\rho p$ final hadron states. For the 
resonant amplitudes, a unitarized Breit-Wigner ansatz that accounts for the transition between the same and different resonances 
in the dressed resonant propagator was employed, which makes the resonant amplitudes consistent with restrictions imposed by a 
general unitarity condition~\cite{Mokeev:2012vsa}. Analysis of the $\gamma_{r,v} p \rightarrow \pi^+\pi^-p$ differential cross sections in the previously unexplored kinematic areas of 1.6 GeV $<$ $W$ $<$ 2.0 GeV (photoproduction) and of 1.4 GeV $<$ $W$ $<$ 2.0 GeV and 2.0 GeV$^2$ $<$ $Q^2$ $<$ 5.0 GeV$^2$ (electroproduction)
reveals no evidence for the contribution of mechanisms other than  those the employed in the previous JM model version \cite{Mokeev:2008iw,Mokeev:2012vsa,Mokeev:2015lda,Golovatch:2018hjk}. However, we modified the parameterization of the extra-contact terms in the $\pi \Delta$ sub-channels and also update the ansatz for the description of the non-resonant contributions in the $\rho p$ and $\pi^+ F_{15}^0(1685)$  sub-channels. After these modifications, a good description of all available $\gamma_{r,v} p \rightarrow \pi^+\pi^-p$ differential and integrated cross sections was achieved in the entire kinematics area covered by the recent CLAS measurements \cite{Golovatch:2018hjk,Isupov:2017lnd,Trivedi:2018rgo} and Table~\ref{clas_data}). The quality of the data description can be seen from the values of $\chi^2/d.p.$ listed in Table~\ref{fit_chi2} and from the range of $\chi^2/d.p.$ $<$ 1.31 for the computed differential cross sections selected in the photoproduction data fit \cite{Golovatch:2018hjk}. A representative example of the differential cross section description together with the resonant and non-resonant contributions inferred from the CLAS $\pi^+\pi^-p$ photoproduction data \cite{Golovatch:2018hjk} is shown Fig.~\ref{res_backgr}. An example of the electroproduction data description is shown in Fig.~\ref{trivedi_fit}. Credible isolation of the resonant contributions achieved in the data fit is vital for the extraction of the resonance photo-/electrocouplings.

\section{Extending Knowledge on the Resonance Spectrum/Structure from CLAS $\pi^+\pi^-p$ Photo-/Electroproduction Data}
\label{results}
The CLAS data for the $\pi^+ \pi^- p$ photoproduction channel play a critical role in the extraction of the photocouplings of higher-lying nucleon excited states ($M >1.60$~GeV), which decay preferentially to the $\pi\pi N$ final states, e.g. $\Delta(1620)1/2^-$,  $\Delta(1700)3/2^-$, $N(1720)3/2^+$, and the $N'(1720)3/2^+$ candidate state. The $\pi^+\pi^-p$ channel has the largest cross section among
the studied $\pi \pi N$ photoproduction channels~\cite{Thoma:2005ca}. The resonance photocouplings extracted from the CLAS data are listed in Table~\ref{nstpar} and compared with the resonance
photocoupling ranges in the PDG2018 \cite{Tanabashi:2018oca} and the results of the multichannel analysis \cite{Sokhoyan:2015fra}. There is good agreement in the magnitude and sign of the photocouplings between our results and the photocoupling
ranges in the PDG listings. On the other hand, for several resonances, the photocouplings determined
from the multichannel analysis are different from ours. Implementation of our $\pi^+\pi^-p$
photoproduction data into the global multichannel analyses will essentially improve our knowledge on the photocouplings and hadronic decay parameters 
of the resonances in the mass range of $W$ $>$ 1.6 GeV.

The preliminary results on the $\gamma_{v}pN^*$ electrocouplings for the $N(1440)1/2^+$ and $N(1520)3/2^-$ resonances have become available at 2.0 GeV$^2$ $<$ $Q^2$ $<$ 5.0 GeV$^2$ from the $\pi^+\pi^-p$ electroproduction channel for the first time. The electrocouplings were determined from a fit of the recent CLAS $\pi^+\pi^-p$ electroproduction data \cite{Isupov:2017lnd,Trivedi:2018rgo} at 1.4 GeV $<$ $W$ $<$ 1.55 GeV and  2.0 GeV$^2$ $<$ $Q^2$ $<$ 5.0 GeV$^2$. Representative example of the computed cross sections selected in the data fit with $\chi^2/d.p.$ $<$ 1.46 are shown in Fig.~\ref{trivedi_fit}. The electrocouplings of the $N(1440)1/2^+$ and $N(1520)3/2^-$ resonances determined at 2.0 GeV$^2$ $<$ $Q^2$ $<$ 5.0 GeV$^2$ from the CLAS $\pi^+\pi^-p$ data \cite{Trivedi:2018rgo} are shown in Fig.~\ref{p11elec} and Fig.~\ref{d13elec} along with the previously available results from CLAS \cite{Mokeev:2012vsa,Mokeev:2015lda,Aznauryan:2009mx} and theoretical expectations \cite{Segovia:2015hra,Aznauryan:2012ec,Aznauryan:2018okk}. Consistent results for the $\gamma_vpN^*$ electrocouplings of the $N(1440)1/2^+$ and $N(1520)3/2^-$ resonances in a wide $Q^2$-range from 0.2 GeV$^2$ to 5.0 GeV$^2$, which have been determined in independent analyses of the dominant meson electroproduction channels $\pi N$ and $\pi^+\pi^-p$, shown in Fig.~\ref{p11elec} and Fig.~\ref{d13elec}, validate the credible extraction of these quantity in a nearly model independent way. This success also demonstrates the capabilty of the JM model for the credible extraction of the $N^*$ parameters from independent studies of $\pi^+\pi^-p$ electroproduction.

Full information on the electrocouplings of most excited nucleon states extracted from the CLAS data can be found in Ref.~\cite{Blin:2019fre} and in the web pages of the Refs. \cite{mokeev-wp,isupov-wp}.

\begin{table}[htb!]
\begin{center}
\caption{Masses and the total and partial hadronic decay widths to the $\pi\Delta$ and $\rho p$ final hadron states of the convention $N(1720)3/2^+$ and candidate $N'(1720)3/2^+$ resonances obtained from the CLAS $\pi^+\pi^-p$ data \cite{Golovatch:2018hjk,Ri03} fit within the JM model.}
\label{decays_conv_miss}      
\begin{tabular}{|c|c|c|}
\hline
Resonance & $N'(1720)3/2^+$ & $N(1720)3/2^+$ \\ \hline
Mass, GeV & 1.715-1.720     & 1.743-1.753    \\ \hline 
Total     & 120 $\pm$ 6     & 112 $\pm$ 8     \\
width, MeV &              &                   \\ \hline 
Branching & 46-64         &    38-55            \\
fraction ($\pi\Delta$), \%  &          &          \\ \hline
Branching &   3-13        & 23-49 \\
fraction ($\rho p$), \%   &             &          \\  \hline
\end{tabular}
\end{center}
\end{table}

The electrocouplings of the $N(1440)1/2^+$ resonance were evaluated 
starting from the QCD Lagrangian within the continuum QCD Dyson-Schwinger Equation (DSE) approach~\cite{Segovia:2015hra}. The DSE predictions are shown in Fig.~\ref{p11elec} by the solid lines. The DSE evaluations are applicable at photon virtualities  where the inner quark core becomes the major contributor to the resonance  structure at $Q^2$ above 1-2 GeV$^2$. In this range of photon 
virtualities, they offer a good description of the experimental results on the $N(1440)1/2^+$ electrocouplings, which was achieved with {\it exactly the same} dressed quark mass function as the one employed in the previous evaluations of the electromagnetic elastic nucleon form factors and the magnetic $p \to \Delta(1232)3/2^+$  form factor~\cite{Segovia:2014aza}. Furthermore, the dressed quark mass function employed in the evaluations of the aforementioned electroexcitation amplitudes  was computed  starting from the QCD Lagrangian \cite{Binosi:2014aea}. This success conclusively demonstrates the capability of gaining insight into the strong QCD dynamics underlying the dominant part of hadron mass generation from the combined experimental results on the nucleon elastic form factors and the $\gamma_vpN^*$ electrocouplings analyzed within continuum QCD approaches. In the near term future the $\Delta(1600)3/2^+$ electrocouplings will become available from the CLAS $\pi^+\pi^-p$ electroproduction data \cite{Isupov:2017lnd,Trivedi:2018rgo}. They will be confronted with the parameter-free prediction from the continuum QCD \cite{Lu:2019bjs} obtained with the same dressed quark mass function as employed in the DSE computations of the nucleon elastic form factors and the $\Delta(1232)3/2^+$ and $N(1440)1/2^+$ electrocouplings. If universality of the dressed quark mass function is confirmed, this will validate the credible access to the hadron mass generation dynamics from the data on nucleon elastic form factors and the electrocouplings of different excited nucleon states.  

\begin{table*}[htb!]
\begin{center}
\caption{\label{hadrp13phel} $N$(1720)3/2$^+$ hadronic decays determined from the independent fits to the data on charged double pion photo- \cite{Golovatch:2018hjk} and electroproduction \cite{Ri03} off protons accounting only for contributions from conventional resonances.}
\begin{tabular}{|c|c|c|c|}
\hline
 Resonance            & $N^*$ total width       & Branching fraction            & Branching fraction         \\
state                & MeV                     &  for decays to $\pi\Delta$    &  for decays to $\rho N$    \\
\hline
 $N$(1720)3/2$^+$       &                         &                               &                            \\
 electroproduction    &  126.0 $\pm$ 14.0       &    64\% - 100\%               & $<$ 5\%                    \\
 photoproduction      &  160.0 $\pm$ 65.0       &   14\% - 60\%                 &  19\% - 69\%               \\
\hline
\end{tabular}
\end{center}
\end{table*}

The novel light front quark model that incorporates the parameterized momentum-dependent quark mass with parameters adjusted to the data on the $Q^2$-evolution of the nucleon elastic form factors \cite{Aznauryan:2018okk,Aznauryan:2012ec} was developed for the description of the $\gamma_vpN^*$ electrocouplings. The model also accounts for the contributions from the meson-baryon cloud. It was found that the implementation of the momentum-dependent dressed quark mass is absolutely needed in order to reproduce the behavior of the nucleon elastic form factors at $Q^2$ $>$ 2.0 GeV$^2$. A successful description of the electrocouplings of many resonances in the mass range up to 1.7~GeV was achieved at $Q^2$ $>$ 2.0 GeV$^2$ with the same momentum- dependent quark mass as that used for the successful description of the nucleon elastic form factors. The description of the $N(1440)1/2^+$ and $N(1520)1/2^+$ electrocouplings is shown in Figs.~\ref{p11elec} and~\ref{d13elec} by the dashed lines. This success within this conceptually different  phenomenological framework compared to the continuum QCD approaches emphasizes the relevance of the dressed quark with running mass as an active component in the structure of the ground and excited nucleon states.

The studies of the combined $\pi^+\pi^-p$ photo- and electroproduction data with CLAS have revealed evidence for the existence of the new candidate state $N'(1720)3/2^+$. The contribution from the $N'(1720)3/2^+$ has been observed, in addition to several other new baryon states, from a global multi-channel analyses of the exclusive meson photoproduction data with decisive impact from CLAS data on $KY$ photoproduction \cite{Bu19n,Anisovich:2017bsk}. For the first time signals from the $N^{\, '}(1720)3/2^+$ candidate state have been observed in analyses of the CLAS  $\gamma_{v}p \rightarrow \pi^+\pi^-p$ electroproduction data \cite{Ri03}, which shows a pronounced structure in the $W$-dependence of the fully integrated cross sections at $W$ $\approx$ 1.7 GeV and in all $Q^2$ bins covered by the measurements. 

Further studies of the CLAS $\gamma_{r,v}p \rightarrow \pi^+\pi^-p$ photo- and electroproduction cross sections \cite{Golovatch:2018hjk,Ri03} have been carried out within the framework of the current version of the JM model outlined in Sec.~\ref{data_anal}. All nine single-differential cross sections were included in the data fits carried out independently for the photo- and electroproduction data. In the fits we simultaneously varied the electrocouplings and decay widths to the $\pi \Delta$ and $\rho N$ final hadron states for all resonances that contribute to the region of the structure at $W \approx$ 1.7 GeV.  We also simultaneously varied the non-resonant parameters of the JM model. 

The hadronic decay widths of each resonance remain the same in all $Q^2$-bins in the electroproduction data fit, while for the photoproduction data, the resonance hadronic couplings were varied independently. We found that for the description of all six angular distributions in the bins of $W$ and $Q^2$ at $W \approx$ 1.7 GeV, an essential contribution from the resonances with spin-parity $J^{\pi}$=3/2$^+$ is needed. Therefore, we compare the two possibilities for description of the $\pi^+\pi^-p$ photo- and electroproduction data at $W \approx$ 1.7 GeV: either by considering the contribution from the conventional $N(1720)3/2^+$ resonance or by implementing in addition a contribution from the new $N'(1720)3/2^+$ state with parameters fit to the CLAS data. Both methods provided a reasonable description of the photoproduction data with $\chi^2/d.p.$ $<$ 1.31 (1.6 GeV $<$ W $<$ 2.0 GeV) accounting for the contribution from the statistical and dependent from the final hadron state kinematic variable systematic data uncertainties, as well as of the electroproduction data with $\chi^2/d.p.$ $<$ 3.0 (1.65 GeV $<$ $W$ $<$ 1.8 GeV) accounting for the statistical data uncertainties only. Representative examples of the description of the photo- and electroproduction differential cross sections are shown in Fig.~\ref{missing_res_sect}. The results in Fig.~\ref{missing_res_sect} were obtained accounting for the contributions from the $N'(1720)3/2^+$ candidate-state, but the fit quality remains almost the same if $N'(1720)3/2^+$ candidate-state is taken out and the photo-/electrocouplings of the conventional $N(1720)3/2^+$ resonance are fit to the data without the contribution from the new candidate state.

On the other hand, as is shown in Table~\ref{hadrp13phel}, the decay widths of the conventional $N(1720)3/2^+$ resonance to the $\pi \Delta$ and $\rho p$ final hadron states inferred from independent analyses of $\pi^+\pi^-p$ photo- and electroproduction depend considerably on the presence or absence of the $N'(1720)3/2^+$ candidate-state. In the case where the contributions from only conventional resonances are taken into account, the $N(1720)3/2^+$ decays into the $\rho N$ final state inferred the data fit on charged double pion photo- and electroproduction off protons differ by more than a factor of four.   Since the resonance hadronic decay widths should be $Q^2$-independent, this makes it impossible to describe both the $\pi^+\pi^-p$ photo- and electroproduction cross sections when only contributions from conventional resonances are taken into account. 

By implementing a new $N^{'}(1720)3/2^+$ baryon state with the parameters inferred from the CLAS $\pi^+\pi^-p$ data fit a successful description of all nine single-differential $\gamma_{r,v}p \rightarrow \pi^+\pi^-p$  photo- and electroproduction cross sections has been achieved. Furthermore, the hadronic decay widths of all resonances in the third resonance region as inferred from the fits at different $Q^2$ remain $Q^2$-independent in the entire range of photon virtualities up to 1.5 GeV$^2$ that is covered by the CLAS measurements \cite{Golovatch:2018hjk,Ri03}. This is a strong evidence for the existence of the new $N'(1720)3/2^+$ baryon state. Indeed, if  the implementation of a new baryon state would just effectively describe the non-resonant contributions, then it would not be possible to reproduce the CLAS $\pi^+\pi^-p$  data in a wide $Q^2$-range with $Q^2$-independent decay widths of such a  fake resonance because of the pronounced evolution of the $\pi^+\pi^-p$ non-resonant parts with $Q^2$. The successful description of the CLAS $\pi^+\pi^-p$ photo- and electroproduction data for 0. GeV$^2$ $<$ $Q^2$ $<$ 1.5 GeV$^2$ with $Q^2$-independent  decay widths of both the $N'(1720)3/2^+$ and $N(1720)3/2^+$ resonances into $\pi\Delta$ and $\rho p$ final hadron states, validates the existence of the new baryon state $N'(1720)3/2^+$ in a nearly model- independent way. 

The results on the $Q^2$-evolution of the $N'(1720)3/2^+$ and $N(1720)3/2^+$ $\gamma_{v}pN^*$ electrocouplings determined from the CLAS data fit are shown in Fig.~\ref{miss_conv_elec}. Their rather different $Q^2$-evolution emphasizes the contributions from both the $N'(1720)3/2^+$ and $N(1720)3/2^+$ excited nucleon states into the $\pi^+\pi^-p$ photo- and electroproduction in the third resonance region. The masses and the total and partial hadronic decay widths of the $N'(1720)3/2^+$ and $N(1720)3/2^+$ resonances are listed in Table~\ref{decays_conv_miss}. The conventional $N(1720)3/2^+$ and candidate $N'(1720)3/2^+$ states have the same spin-parities. Their masses are slightly differenrt and the total decay widths are almost the same within the data uncertainties. However, the different patterns for their decays to the $\pi\Delta$ and $\rho p$ final hadron states and different $Q^2$-evolution of their electrocouplings allow us to identify these two different resonances. So far, the $N'(1720)3/2^+$ is the only ``missing" baryon state for which the experimental results on the $Q^2$-evolution of the electroexcitation amplitudes have become available. These results offer insight into the structure of the so-called ``missing" baryon states, paving a way towards understanding the peculiar features of these resonances which have made them so elusive from the observation in experiments for a long time. In the studies of the conventional and exotic baryons within the chiral quark-soliton model \cite{Yang:2018gju}, the new baryon state $N(1726)3/2^+$ was predicted as a member of the 27-plet. The mass and spin-parity of this predicted state are consistent with the observed $N'(1720)3/2^+$ candidate-resonance. When the predictions from this or other quark models on the $N'(1720)3/2^+$ electrocouplings will become available, they will shed light on the nature of the new baryon state observed in the CLAS $\pi^+\pi^-p$ photo- and electroproduction data. This success also demonstrates the importance of combined studies of exclusive photo- and electroproduction, both for the validation of new baryon states and for the further search for new excited nucleon states.

\section{Conclusion and Outlook}
\label{summary}
Studies of exclusive $\pi^+\pi^-p$ photo- and electroproduction with CLAS have extended the information on the nucleon resonance photo- and electroexcitation amplitudes. The photocouplings of almost all nucleon resonances in the mass range from 1.6 GeV to 2.0 GeV have become available for the first time from the studies of $\pi^+\pi^-p$ photoproduction. The entries on the photocouplings and $\pi\Delta$ and $\rho p$ hadronic decays of most resonances in the mass range from 1.6 GeV to 2.0 GeV have been updated in the PDG by results from CLAS \cite{pdg}.

Studies of the combined $\pi^+\pi^-p$ photo- and electroproduction data have provided evidence for the candidate-state $N'(1720)3/2^+$. The contribution from this state is needed in order to describe both photo- and electroproduction data with $Q^2$-independent masses and total and partial decay widths of the $N(1720)3/2^+$-conventional and $N'(1720)3/2^+$ candidate resonances to the $\pi \Delta$ and $\rho p$ final hadron states. The $N'(1720)3/2^+$ is the only ``missing" baryon state observed so far for which the information on the $Q^2$-evolution of the electroexcitation amplitudes has become available.  These results allow us to gain insight into the structure of ``missing" resonances and eventually to understand why these states have remain elusive from the observation  for a so long time. Predictions on the $Q^2$-evolution of the ``missing" resonance electrocouplings from different models is the urgently needed next step towards approaching this challenging objective.

The $\gamma_{v}pN^*$ electrocouplings have become available from the CLAS $\pi^+\pi^-p$ electroproduction data at 0.2 GeV$^2$ $<$ $Q^2$ $<$ 5.0 GeV$^2$ for the resonances in the mass range up to 1.6 GeV and at 0.5 GeV$^2$ $<$ $Q^2$ $<$ 1.5 GeV$^2$ for most resonances in the mass range from 1.6 GeV to 1.8 GeV.
Consistent results on nucleon resonance electrocouplings from independent studies of the $N\pi$ and $\pi^+\pi^-p$ electroproduction channels with different non-resonant contributions validate the credible extraction of these quantities. Physics analyses of the CLAS results on the $\gamma_{v}pN^*$ electrocouplings within the framework of the continuum QCD approach \cite{Craig19n,Segovia:2014aza,Segovia:2015hra} and the novel relativistic quark model \cite{Aznauryan:2018okk} have revealed the dressed quarks with momentum-dependent masses  as active degrees of freedom in $N^*$ structure. The CLAS results on the $\gamma_{v}pN^*$ electrocouplings of different resonances are of particular importance in order to gain insight into the strong QCD dynamics underlying the generation of the dominant part of hadron mass encoded in the momentum dependence of running dressed quark mass. Consistent results on the dressed quark mass function from independent studies of excited nucleon states of different structure validate credible access to this key ingredient of strong QCD.

In the future, the electrocouplings of most excited nucleon states in the mass range up to 2.0 GeV will become available at 2.0 GeV$^2$ $<$ $Q^2$ $<$ 5.0 GeV$^2$. They will be obtained from the recent CLAS $\pi^+\pi^-p$ electroproduction data \cite{Isupov:2017lnd,Trivedi:2018rgo} analyzed within the framework of the updated JM model outlined in Section~\ref{data_anal}. This essential increase of the information on the $\gamma_{v}pN^*$ electrocouplings will open up new avenues for the exploration of many facets of strong QCD in generation of the excited nucleon spectrum. The insight into the structure of most $N^*$ will allow us to address the challenging open problems of the Standard Model in the strong QCD regime, including: a) insight into the dressed quark mass generation in different resonances in connection with hadron mass generation; b) studies of the dressed-quark-gluon vertex and di-quark correlation complexity in the structure of orbital excited resonances; c) insight into dynamical chiral symmetry breaking  from the data on electrocouplings of the resonances parity-partners; d) search for new states of baryon matter and exploration of their structure. Furthermore, the information on electrocouplings of most excited nucleon states makes it possible to determine the resonance contributions in inclusive electron scattering \cite{Blin:2019fre,Blin19n} and  gain insight into the behavior of parton distributions in the ground state nucleons at close to unity values of Bjorken variable $x_{B}$ bridging the efforts on the nucleon structure studies in the resonance and deep inelastic scattering regions.

\begin{acknowledgement} 
This work was supported by the U.S. Department of Energy, Office of Science, Office of Nuclear Physics under contract DE-AC05-06OR23177
\end{acknowledgement}


\end{document}